\documentclass{article}
\usepackage{spconf,amsmath,graphicx,amssymb}
  \usepackage{subcaption}
  \usepackage{bm}
  \usepackage{multirow}
  \usepackage{booktabs}
  \usepackage{textgreek}
  \usepackage{gensymb}
  \usepackage{color,soul}

\title{Location-based training for multi-channel talker-independent speaker separation}
\name{Hassan Taherian$^1$, Ke Tan$^1$,  {\normalfont{and}} DeLiang Wang$^{1,2}$ \thanks{This research was supported in part by a National Science Foundation grant (ECCS-1808932), the Ohio Supercomputer Center, and the Pittsburgh Supercomputer Center (NSF ACI-1928147).}}
\address{
$^1$Department of Computer Science and Engineering, The Ohio State University, USA\\
$^2$Center for Cognitive and Brain Sciences, The Ohio State University, USA\\	
\texttt{\small\{taherian.1, tan.650\}@osu.edu, dwang@cse.ohio-state.edu}}
\begin{document}
\maketitle

\begin{abstract} 
	Permutation-invariant training (PIT) is a dominant approach for addressing the permutation ambiguity problem in talker-independent speaker separation. Leveraging spatial information afforded by microphone arrays, we propose a new training approach to resolving permutation ambiguities for multi-channel speaker separation. The proposed approach, named location-based training (LBT), assigns speakers on the basis of their spatial locations. This training strategy is easy to apply, and organizes speakers according to their positions in physical space. Specifically, this study investigates azimuth angles and source distances for location-based training. Evaluation results on separating two- and three-speaker mixtures show that azimuth-based training consistently outperforms PIT, and distance-based training further improves the separation performance when speaker azimuths are close. Furthermore, we dynamically select azimuth-based or distance-based training by estimating the azimuths of separated speakers, which further improves separation performance. LBT has a linear training complexity with respect to the number of speakers, as opposed to the factorial complexity of PIT. We further demonstrate the effectiveness of LBT for the separation of four and five concurrent speakers.

\end{abstract}

\section{Introduction}
	\label{sec:intro}
	\vspace{-.5em}

Recent speaker separation methods based on deep neural networks (DNNs) have substantially improved separation performance~\cite{zeghidour2021wavesplit, luo2020dual, wang2021multi}. To train a talker-independent separation model, where test speakers can be different from training ones, each output layer of a DNN model needs to be associated with one distinct speaker in the mixture~\cite{wang2018supervised}. Ambiguity in speaker assignment would lead to conflicting gradients during training. This permutation ambiguity problem also arises in DNN based speaker diarization~\cite{Fujita2019Interspeech} and multi-source speaker localization~\cite{subramanian2021deep}. Main solutions to this problem include deep clustering \cite{hershey2016deep} and permutation invariant training (PIT)~\cite{kolbaek2017multitalker}. 
In deep clustering, a DNN maps time-frequency units to embedding vectors with an objective function that is invariant to speaker permutations. These embedding vectors are then clustered via the K-means algorithm to estimate the ideal binary mask.
On the other hand, PIT resolves the permutation ambiguity
by examining the losses from all possible output-speaker permutations, and it does not require an additional clustering step.

Deep clustering and PIT were originally developed for monaural speaker separation.
The availability of multi-channel recordings provides a spatial dimension, which is missing in monaural recordings.  
We believe that the permutation ambiguity problem can be naturally avoided by leveraging spatial relations of multiple speakers. It is a basic fact that multiple speakers cannot occupy the same spatial location. In this study, we propose a new training approach to achieving multi-channel talker-independent speaker separation. To resolve the permutation ambiguity problem, we propose location-based training (LBT), which assigns DNN output layers according to speaker locations. Specifically, we investigate azimuth-based and distance-based training, which makes assignments based on speaker azimuth angles and distances relative to a microphone array. 

Our separation model uses multi-channel complex spectral mapping (MC-CSM)~\cite{wang2020multi}.  Evaluation results show that azimuth-based training outperforms PIT in both anechoic and reverberant environments, while distance-based training is more superior in conditions where  speakers have close azimuths. To combine the relative advantages of azimuth-based and distance-based training, we dynamically select the two training criteria on the basis of azimuth estimates of separated speakers. In this case, speaker localization is performed by mask-weighted generalized cross-correlation with phase transform (GCC-PHAT)~\cite{8492455}. 

In contrast to PIT whose training complexity is factorial to the number of speakers, LBT has a linear computational complexity. Given the low complexity of LBT, multi-channel DNN models can be trained efficiently for a large number of concurrent speakers. Moreover, separated speakers are naturally ordered according to their spatial locations. This facilitates the integration of a speaker separation model with downstream speech processing tasks such as speaker localization, diarization, recognition, and automatic speech recognition  \cite{wang2021multi, yoshioka2019advances, taherian2021time}.

This work expands our preliminary study \cite{ke_thesis} which illustrates the effectiveness of azimuth-based training for two-speaker mixtures in anechoic conditions. A recent study also considers ordering speakers based on azimuth angles  for the task of multi-source speaker localization~\cite{subramanian2021deep}.

\section{System Description}
\vspace{-.5em}
	\label{sec:sys_description}
\subsection{Location-based training}
\vspace{-.5em}
A primary approach to talker-independent speaker separation utilizes utterance-level PIT to address the permutation ambiguity problem~\cite{luo2020dual,wang2021multi, taherian2021time}. Utterance-level PIT uses fixed output-speaker pairings for a whole utterance, and selects the optimal pairing that minimizes the loss function over all possible speaker permutations~\cite{kolbaek2017multitalker}:
\begin{equation}
	\mathcal{L}_{\text{PIT}}  = \min_{\phi_{1}, \dots, \phi_{N} \in \Phi}  \sum_{n=1}^{N} \mathcal{L}(\hat{S}_n, S_{\phi_{n}}),
	\label{eq_pit_loss}
 \end{equation}
where $\hat{S}$ and $S$ are estimated and clean speech signals in the short-time Fourier transform (STFT) domain, respectively. $\mathcal{L}$ denotes a loss function, and symbol $\Phi$ is the set of all permutations of $N$ speakers. 

With the assumption that speakers are still, we propose to utilize the spatial locations of speakers to resolve the permutation ambiguity problem for multi-channel talker-independent speaker separation. In this study, we explore LBT based on speaker azimuth angles and distances relative to the center of a microphone array. 

Let $\theta_{1}, \theta_{2},\dots \theta_{N} \in [0, 2\pi)$ be the sorted speaker azimuths relative to the microphone array. The loss of azimuth-based training is defined as:
\begin{equation}
	\mathcal{L}_{\text{Azimuth}}  = \sum_{n=1}^{N} \mathcal{L}(\hat{S}_n, S_{\theta_{n}}).
	  \label{eq_azi_loss}
  \end{equation}

\begin{figure}[t]
	\includegraphics[width=0.29\textwidth]{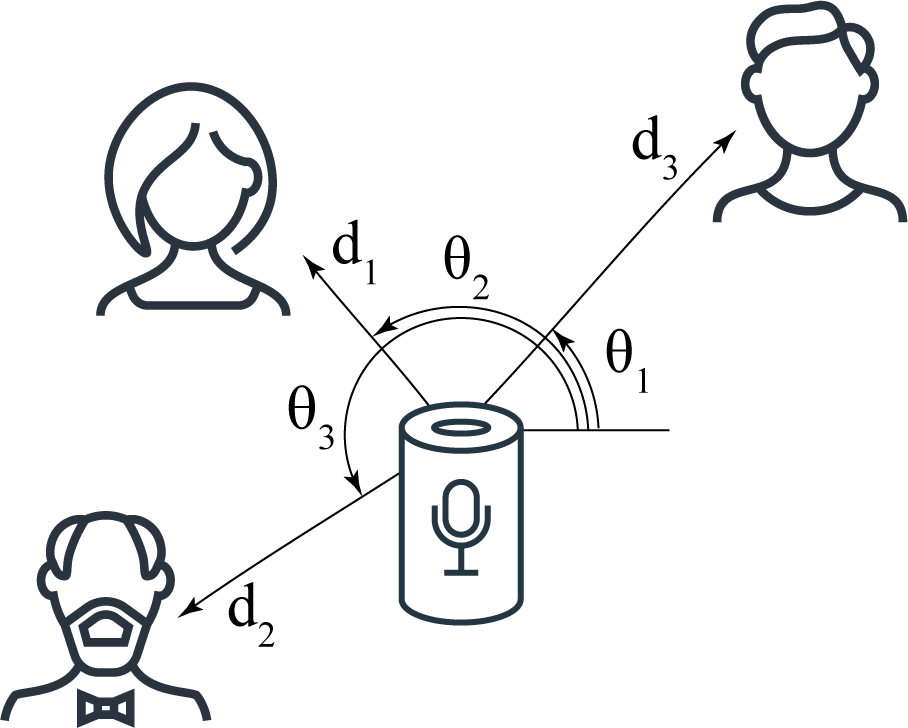}
	\centering
	\vspace{-.6em}
	\caption{\small{Illustration of new training criteria based on speaker azimuths and distances relative to a microphone array. }}
	\vspace{-.9em}
	\label{fig:sketch}
\end{figure}

  Fig.~\ref{fig:sketch} illustrates LBT with 3 speakers. In the case of azimuth-based training output-speaker assignments follow the azimuths order, where the first output is tied to the speaker with the smallest azimuth and the last output is tied to the speaker with the largest azimuth. Note that the azimuth range is dependent on the array geometry. For linear arrays, the azimuth range should be in $[0, \pi)$, due to the well-documented front-back confusion of linear arrays. Similarly, we formulate distance-based training as: 
  \begin{equation}
	\mathcal{L}_{\text{Distance}}  = \sum_{n=1}^{N} \mathcal{L}(\hat{S}_n, S_{d_{n}}),
	  \label{eq_dist_loss}
  \end{equation}
where $d_{1}, d_{2},\dots, d_{N}$ are speaker distances to the microphone array in an ascending order. With this criterion, we assign the nearest speaker to the first output layer and the farthest speaker to the last output layer (see Fig.~\ref{fig:sketch}).

In both criteria, output-speaker assignments are based on the relative positions of  speakers. By leveraging spatio-temporal patterns in the multi-channel input, LBT resolves the permutation ambiguity problem through the consistent pairings of DNN output layers and speaker locations.

\subsection{Multi-channel complex spectral mapping}
\vspace{-.5em}

Assuming a fixed array geometry, MC-CSM estimates the complex spectrogram of target speech received at the reference microphone from that of the multi-channel noisy mixture. It is shown that MC-CSM can implicitly learn the spectral and spatial information within the array signals~\cite{wang2020multi}. We employ the Dense-UNet architecture proposed in \cite{liu2019divide} for single- and multi-channel complex spectral mapping. We stack the real and imaginary components of the mixture STFT at all microphones and pass them into the Dense-UNet as input. The model estimates the cIRM (complex ideal ratio mask)~\cite{williamson2016complex}, which is then multiplied by the complex spectrogram of the input mixture at the reference microphone.

The Dense-UNet architecture includes 4 downsampling and upsampling layers interleaved with 9 densely-connected convolutional blocks. Each dense block contains 5 convolutional layers, each of which has 64 channels, a kernel size of $3 \times 3$ and a stride of $1 \times  1$. The middle layer in each dense block is replaced with a frequency mapping layer  to deal with inconsistencies between different frequency bands~\cite{liu2019divide}. We adopt the loss function in~\cite{wang2020multi} for each output-speaker pair, which is based on $\ell_1$ norm of real and imaginary spectrograms of estimated and target speech with an additional magnitude loss term:
\begin{equation}
	\mathcal{L}_{\text{RI+Mag}}(\hat{S}, S) =  \mathcal{L}_{\text{RI}}(\hat{S}, S) + \left\lVert |\hat{S}| - |S| \right\rVert_{1},
\end{equation}
where $|S|$ and $|\hat{S}|$ represent the target and estimated magnitude spectrograms, and $|\hat{S}|$ is calculated from the estimated real and imaginary components $\hat{S}^{(r)}$ and $\hat{S}^{(i)}$:
\begin{equation}
|\hat{S}| = \sqrt{(\hat{S}^{(r)})^2 + (\hat{S}^{(i)})^2 }.
\end{equation}
In addition,
\begin{equation}
	\mathcal{L}_{\text{RI}}(\hat{S}, S) =  \left\lVert \hat{S}^{(r)} - S^{(r)} \right\rVert_{1} + \left\lVert \hat{S}^{(i)} - S^{(i)} \right\rVert_{1}.
\end{equation}


	
\section{Experimental Setup}
\vspace{-.5em}
	\label{sec:exp_setup}
Our experiments use simulated room impulse responses (RIRs) for evaluation. We generate RIRs for a 7-channel circular microphone array using the image method~\cite{allen1979image, scheibler2018pyroomacoustics}. The microphone array comprises 6 microphones uniformly distributed on a circle with a radius of 4.25~cm and one microphone at the center of the circle. We simulate rectangular rooms with random length, width and height dimensions in the range of [4$\times$4$\times$3, 6$\times$6$\times$4] meters, with the microphone array placed in the center of the room. 

The speech sources are placed in positions randomly selected from 72 candidate azimuth positions in the range of -180$^\circ$ to 180$^\circ$ with a 5$^\circ$ resolution. For a speaker pair $(i,j)$, the source-array distances $d_i$ and $d_j$ are randomly selected such that $|d_i - d_j| \geqslant 0.2$ m. Moreover, the minimum source-array distance is set to 0.3~m. We assume that speech sources are placed at the same height as the microphone array. 

We create speech mixtures with 2 and 3 speakers in both anechoic and reverberant conditions. The multi-channel mixtures are created by spatializing the WSJ0-2mix and WSJ0-3mix datasets~\cite{hershey2016deep} with the simulated RIRs, which include 20000, 5000 and 3000 mixtures in the training, validation and test sets, respectively. For the reverberant mixtures, the reverberation time (T60) is randomly sampled between 0.15 and 0.6 seconds. Note that we treat the center microphone as the reference microphone. For all speakers, the direct-path (anechoic) signal at the reference microphone is used as the target signal. All signals are sampled at 16~kHz.

\section{Evaluation Results}
\vspace{-.5em}
	\label{sec:eval}
	\begin{table}   
		\caption{\small{ESTOI (\%), PESQ, SI-SNR (dB) and SDR (dB) of
		different training criteria on reverberant 2-speaker and 3-speaker mixtures with 5$^\circ$ and 1$^\circ$ resolutions of azimuth spacing. `Combined' refers to the combination of azimuth-based and distance-based training.}}
		\vspace{-.5em}
		\label{tab:rev}
		\centering
		\resizebox{0.484\textwidth}{!}{
			\renewcommand{\arraystretch}{1.30}
		  \begin{tabular}{{l l c  cc cc }}
		  \toprule
		  \multicolumn{2}{c}{}  & Criterion & ESTOI & PESQ & SI-SNR & SDR     \\  
		  \midrule
		  \multirow{5}{*}{\rotatebox[origin=c]{90}{\textbf{2-speaker~/~5$^\circ$}}} 
		  &Unprocessed &--& 37.42&	1.61&	-8.15&	-1.72\\ 
		  &SC-CSM   &PIT& 63.44&	2.27&	-0.28&	3.77\\ 
		  &MC-CSM  &PIT& 	78.47&	2.90&	5.71&	8.94\\
		  &MC-CSM   &Azimuth& \textbf{82.07}&	\textbf{3.06}&	\textbf{7.01}&	\textbf{9.91}\\
		  &MC-CSM    &Distance& 80.01&	2.96&	6.45&	9.07\\
		  \midrule 
		  \multirow{5}{*}{\rotatebox[origin=c]{90}{\textbf{3-speaker~/~5$^\circ$}}} 
		  &Unprocessed &--& 27.78&	1.36&	-9.49&	-4.64\\  
		  &SC-CSM   &PIT& 42.92&	1.62&	-3.83&	0.57\\ 
		  &MC-CSM &PIT& 67.34&	2.46&	4.40&	6.97\\  
		  &MC-CSM &Azimuth& \textbf{69.21}&	\textbf{2.58}&	\textbf{4.79}&	\textbf{7.92}\\  
		  &MC-CSM &Distance&  66.63&	2.41&	4.24&	6.62\\

		  \midrule 
		  \multirow{5}{*}{\rotatebox[origin=c]{90}{\textbf{2-speaker~/~1$^\circ$}}} 
		  &Unprocessed &--& 37.36&	1.61&	-8.15&	-1.75\\
		  &MC-CSM  &PIT& 74.78&	2.74&	4.64&	7.88\\ 
		  &MC-CSM   &Azimuth& 80.98&	3.03&	6.66&	9.74\\
		  &MC-CSM    &Distance& 79.75&	2.95&	6.43&	9.13\\
		  &Combined &--& \textbf{81.33}&	\textbf{3.04}&	\textbf{6.76}&	\textbf{9.84}\\ 

		  \bottomrule
		  \end{tabular}
		  }
	\end{table}

We report the results in terms of signal-to-distortion ratio (SDR), scale-invariant signal-to-noise ratio (SI-SNR), perceptual evaluation of speech quality (PESQ), and extended short-time objective intelligibility (ESTOI). As a comparison baseline, we also report the results for the PIT-based single-channel CSM (SC-CSM).
 
Table~\ref{tab:rev} presents the MC-CSM results with different training criteria in the reverberant condition. The first two rows give results with the 5$^\circ$ azimuth resolution.
Regardless of the training criterion, MC-CSM leads to significant performance improvement compared to SC-CSM on two-speaker and three-speaker mixtures. We observe that MC-CSM with azimuth-based training outperforms PIT in all metrics. Although distance-based training underperforms azimuth-based training, it yields comparable results to PIT.
  
\begin{figure}[t]
		\includegraphics[width=0.48\textwidth]{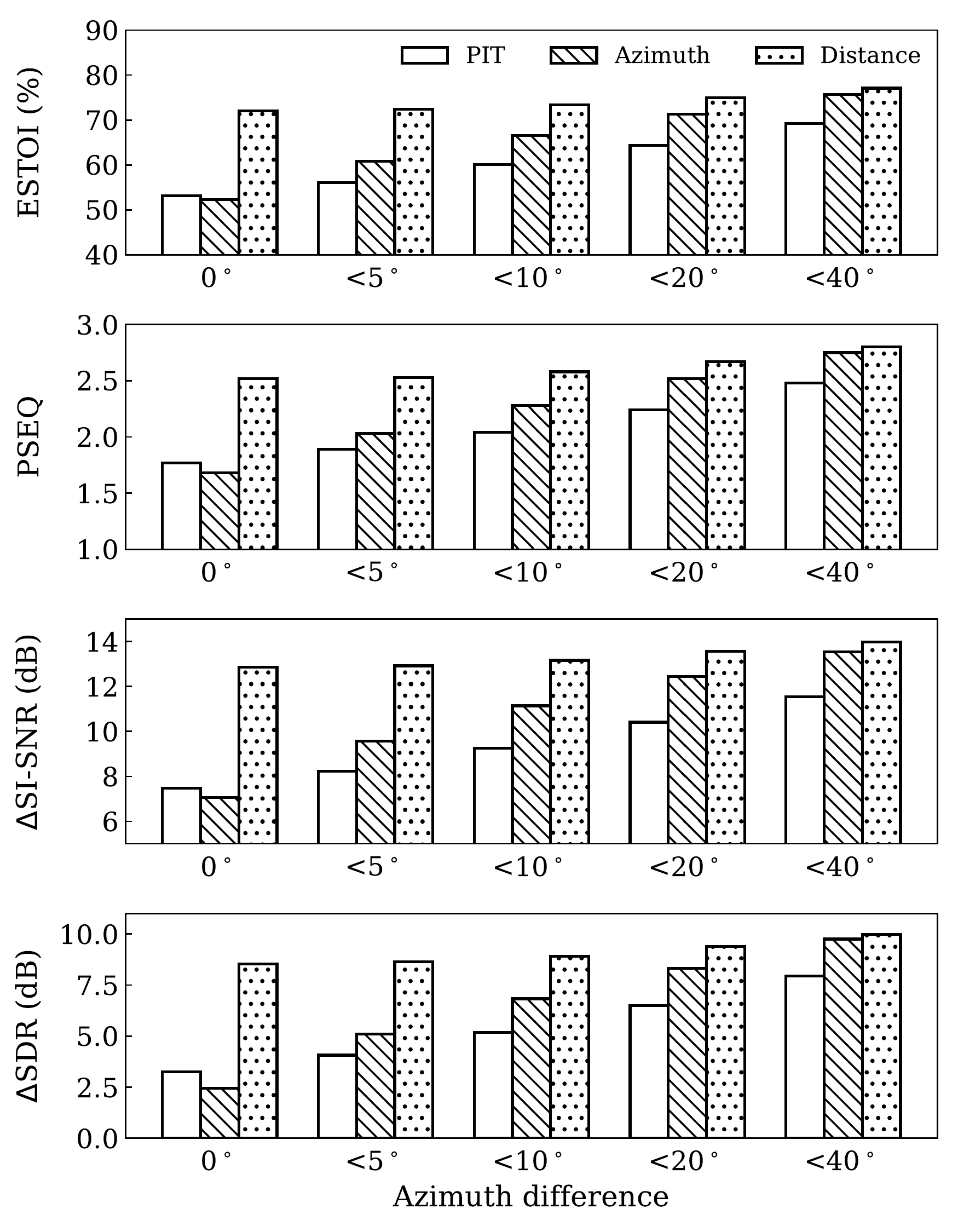}
		\centering
		\vspace{-.5em}
		\caption{\small{
		ESTOI, PESQ, SI-SNR improvement ($\Delta$SI-SNR) and SDR improvement ($\Delta$SDR) of different training criteria with constraint azimuth difference. The models are trained on reverberant WSJ-2mix with 1$^\circ$ resolution.}}
		\label{plot:wsj2mix_difficult}
		\vspace{-.5em}
	\end{figure}

 We additionally train the MC-CSM model on reverberant two-speaker mixtures with a 1$^\circ$ resolution of azimuth spacing. As shown in  the third row of Table~\ref{tab:rev}, similar trends occur for location-based MC-CSM with this finer spatial resolution. To further investigate the effect of LBT, we evaluate MC-CSM on different sets of reverberant two-speaker mixtures where the difference between speaker azimuths is constrained. The results are shown in Fig.~\ref{plot:wsj2mix_difficult}. We observe that the performance of the models with PIT and azimuth-based training significantly degrades when azimuth differences are small. Not surprisingly, distance-based training is relatively insensitive to azimuths and outperforms the other two methods.
  	
To take advantage of both azimuth-based and distance-based training, we perform source localization to estimate the speaker azimuths, and use these estimates to dynamically select between the two LBT criteria. The azimuth of speaker $k$ can be well estimated from a speech mixture using mask-weighted GCC-PHAT \cite{8492455, 1162830}:
\begin{equation}
 \operatorname*{argmax}_{\tau  }\sum_{(p,q) \in \Omega} \sum_{t,f}  \lambda_{k} \text{GCC}_{p,q}(t,f,\tau ), 
\label{eq_masked_gccphat}
\end{equation}
where $\text{GCC}_{p,q}(t,f,\cdot)$ represents the GCC-PHAT function for microphone pair $(p,q)$ at time $t$ and frequency $f$. Symbol $\tau$ denotes the time delay corresponding to a candidate azimuth, and $\Omega$ is the set of  all microphones pairs. Moreover, $\lambda_{k}$ is a ratio mask for speaker $k$, computed using the mixture STFT $Y_{\text{ref}}$ at the reference microphone:
\begin{equation}
	\lambda_{k} = \frac{|\hat{S}_k|^2}{|\hat{S}_k|^2 + |Y_{\text{ref}} - \hat{S}_k|^2}.
\end{equation}
For the 1$^\circ$ resolution experiment, we use an empirical threshold of 20$^\circ$ for dynamic criterion selection. Specifically, we select azimuth-based training if the computed azimuth difference is larger than 20$^\circ$, and distance-based training otherwise. As shown in Table~\ref{tab:rev}, such a combination further improves the results. Note that only 12\% of mixtures in this test set contain speakers with an azimuth difference less than 20$^\circ$.

We present the evaluation results in the anechoic condition in Table~\ref{tab:wsj2mix}. Similar to Table~\ref{tab:rev}, azimuth-based training achieves superior performance to PIT for two- and three-speaker mixtures. However, the performance of distance-based training significantly degrades in the anechoic condition. We conjecture that distance-based training implicitly leverages  direct-to-reverberant ratios (DRRs) from different speakers for speaker separation. The DRR is inversely proportional to the square of the source-microphone distance in reverberant environments~\cite{blauert1997spatial, 8878108}. As the source-microphone distance increases, the energy of the direct sound decreases while the energy of the reverberant sounds remains roughly constant. In the reverberant condition, the model trained with the distance criterion may learn to assign the speaker with the highest DRR to the first output layer and the second highest DRR to the second output layer, and so on. In an anechoic room, the DRR is infinite and thus cannot serve as a discriminative cue to separate between nearer and farther speakers. However, we note that anechoic conditions do not occur in the real world.

  \begin{table}   
	\caption{\small{Comparison of different training criteria for mixtures with the various number of speakers in the anechoic condition.}}
	\vspace{-.5em}
	\label{tab:wsj2mix}
	\centering
	\resizebox{0.484\textwidth}{!}{
	\renewcommand{\arraystretch}{1.35}
	  \begin{tabular}{{l l c  cc cc }}
	  \toprule
	  \multicolumn{2}{c}{}  & Criterion & ESTOI & PESQ & SI-SNR & SDR     \\  
	  \midrule
	  \multirow{5}{*}{\rotatebox[origin=c]{90}{\textbf{2-speaker}}} 
	  &Unprocessed &--& 56.11&	1.89&	0.00&	0.13\\  
	  &SC-CSM   &PIT& 83.01&	2.88&	11.12&	11.55\\
	  &MC-CSM &PIT& 97.60&	4.03&	24.51&	25.09\\ 
	  &MC-CSM  &Azimuth& \textbf{98.49}&	\textbf{4.09}&	\textbf{26.10}&	\textbf{26.68}\\
	  &MC-CSM  &Distance& 82.47&	2.97&	10.77&	11.30\\ 
	  
	  \midrule 
	  \multirow{5}{*}{\rotatebox[origin=c]{90}{\textbf{3-speaker}}} 
	  &Unprocessed &--& 38.54&	1.48&	-4.43&	-4.12\\  
	  &SC-CSM   &PIT& 60.75&	2.05&	4.33&	5.10\\ 
	  &MC-CSM &PIT& 85.10&	3.21&	14.01&	14.58\\ 
	  &MC-CSM &Azimuth& \textbf{90.82}&	\textbf{3.51}&	\textbf{17.13}&	\textbf{17.67}\\
	  &MC-CSM &Distance& 69.06&	2.45&	7.27&	8.00\\ 

	  \midrule 
	  \multirow{3}{*}{\rotatebox[origin=c]{90}{\textbf{4-speaker}}} 
	  &Unprocessed &--&29.36&	1.31&	-7.03&	-6.51\\ 
	  &MC-CSM &PIT& 70.61&	2.64&	8.19&	9.04\\ 
	  &MC-CSM &Azimuth& \textbf{81.48}&	\textbf{3.02}&	\textbf{11.76}&	\textbf{12.41}\\ 

	  \midrule 
	  \multirow{3}{*}{\rotatebox[origin=c]{90}{\textbf{5-speaker}}} 
	  &Unprocessed &--& 23.94&	1.22&	-8.72&	-8.06\\ 
	  &MC-CSM &PIT& 61.39&	2.30&	4.73&	5.88\\ 
	  &MC-CSM &Azimuth& \textbf{70.11}&	\textbf{2.56}&	\textbf{7.24}&	\textbf{8.07}\\

	  \bottomrule
	  \end{tabular}
	  }
\end{table}

We have also evaluated azimuth-based training with four- and five-speaker mixtures. The same simulation procedure for the anechoic condition is used to generate a spatialized version of the WSJ0-4mix and WSJ0-5mix datasets~\cite{nachmani2020voice}. Azimuth-based training outperforms PIT in both four- and five-speaker mixtures.
In addition, as mentioned earlier, the training complexity advantage of LBT over PIT is a lot more evident for such mixtures.
The results suggest that LBT can be potentially used for end-to-end diarization with a large number of speakers~\cite{Fujita2019Interspeech}.
\vspace{-.5em}
\section{Concluding Remarks}
\vspace{-.5em}
	\label{sec:conclude}

We have proposed location-based training as a new training approach for multi-channel talker-independent speaker separation.
We have developed two new training criteria based on speaker azimuth angles and distances to resolve the permutation ambiguity problem. In addition, azimuth-based and distance-based training can be combined to further improve separation performance. LBT outperforms PIT in separation performance as well as training complexity. Future work will extend LBT to separate moving speakers, and nonspeech sources.

\bibliographystyle{IEEEtran}
{\small\bibliography{mybib}}
\end{document}